\documentstyle[preprint,aps]{revtex}
\newcommand{\GG}{$\hat{\Gamma}\hat{\Gamma}^{*}\,$}
\newcommand{\G}{$\hat{\Gamma}(\hat{\Gamma}^{*})\,$}

\newcommand{\bce}{\begin{center}}
\newcommand{\ece}{\end{center}}
\newcommand{\beq}{\begin{equation}}
\newcommand{\eeq}{\end{equation}}
\newcommand{\bea}{\vspace{0.25cm}\begin{eqnarray}}
\newcommand{\eea}{\end{eqnarray}}

\newcommand{\ba}{\begin{array}}
\newcommand{\ea}{\end{array}}

\def\lsim{\mathrel{\rlap{\lower4pt\hbox{\hskip1pt$\sim$}}
    \raise1pt\hbox{$<$}}}	  
\def\gsim{\mathrel{\rlap{\lower4pt\hbox{\hskip1pt$\sim$}}
    \raise1pt\hbox{$>$}}}	  

\def\beq{\begin{equation}}
\def\endeq{\end{equation}}
\def\bea{\begin{eqnarray}}
\def\arr{\begin{eqnarray}}
\def\endarr{\end{eqnarray}}

\makeindex
\begin{document}
\draft
\title{On the missing momentum dependence of the color
transparency effects in $(e,e'p)$ scattering}
\author{O.Benhar$^{1}$,
 S.Fantoni$^{2,3}$,
N.N.Nikolaev$^{4,5}$,
J.Speth$^{4}$,
A.A.Usmani$^{2}$,
B.G.Zakharov$^{5}$
}
\address{
$^{1}$INFN, Sezione Sanit\`{a}, Physics Laboratory,
 Istituto Superiore di Sanit\`{a}. I-00161 Roma, Italy \\
$^{2}$Interdisciplinary Laboratory, SISSA, INFN,
Sezione di Trieste. I-34014, Trieste, Italy \\
$^{3}$International Centre for Theoretical Physics,
Strada Costiera 11, I-34014, Trieste, Italy\\
$^{4}$IKP(Theorie), Forschungszentrum  J\"ulich GmbH.\\
D-52425 J\"ulich, Germany \\
$^{5}$L.D.Landau Institute for Theoretical Physics, \\
GSP-1, 117940, ul.Kosygina 2., V-334 Moscow, Russia }

\date{\today}
\maketitle
\begin{abstract}
We explore
the missing momentum dependence of the CT effects in
quasielastic $(e,e'p)$
scattering.
We develop the coupled-channel multiple-scattering theory (CCMST)
description of final-state interaction including both the coherent
and incoherent rescatterings of the ejectile state.
We demonstrate that the contribution of the off-diagonal
incoherent rescattering does not vanish at low $Q^{2}$ which
is a novel correction to the conventional Glauber theory
evaluation of nuclear transparency. We comment on the
nontrivial impact of this correction on the onset of CT.
The sensitivity of the onset of CT to the
$3q$-nucleon reggeon amplitudes is discussed for the first time.
We present numerical results for nuclear
transparency as a function of the missing momentum
for exclusive $(e,e'p)$ reaction in the
kinematical region of
$Q^{2} \mathrel{\rlap{\lower4pt\hbox{\hskip1pt$\sim$}}
 \raise1pt\hbox{$<$}}  40$ GeV$^{2}$
and
$p_{m} \mathrel{\rlap{\lower4pt\hbox{\hskip1pt$\sim$}}
 \raise1pt\hbox{$<$}} 250$ MeV/c.
Our evaluations
show that at $Q^{2}\sim 10$ CT effects are substantial
only in antiparallel kinematics
at $p_{m,z}\sim -250$ MeV/c. The effect is enhanced
on light nuclei and could be observed in a
high precision experiment.
\end{abstract}

\section{Introduction}
The quasielastic $(e,e'p)$ reaction plays an important
role in the nuclear physics as a tool for investigation of
the nuclear structure. At high $Q^{2}$
it becomes interesting from the point of view of
the particle physics as well.
The perturbative QCD \cite{Brodsky1} suggests that
the dominant mechanism
of $ep$ scattering at high $Q^{2}$ is
an interaction of the virtual photon with  small-size
($\rho\sim 1/Q$) $3q$ configurations in the proton wave
function. It is expected \cite{Mueller1,Brodsky2} that this
mechanism should manifest itself
through the vanishing of final state interaction (FSI)
in $(e,e'p)$ reaction in the limit of high $Q^{2}$
because the small-size $3q$ ejectile state formed
after absorption of the virtual photon will weakly
interact with the spectator nucleons. As a consequence, at high $Q^{2}$
the nuclear transparency, $T_{A}$, defined as a ratio
of the experimentally measured cross section to the theoretical
cross section calculated in the plane wave impulse approximation
(PWIA) should tend to unity, and the experimental missing
momentum distribution be close to the single particle momentum
distribution (SPMD) of the proton in the target nucleus.
The observation of this effect, which is usually referred to
as the color transparency (CT) phenomenon, would be
a direct test of the space-time picture of hard processes
predicted by perturbative QCD.

An accurate evaluation of FSI effects requires a
quantum mechanical treatment of the evolution of the $3q$
ejectile wave function in the nuclear medium.
At large $Q^{2}$
($\gsim 2$ GeV$^{2}$) the kinetic energy of the struck proton
$T_{kin}\approx Q^{2}/2m_{p}$ (here $m_{p}$ is the proton mass)
becomes large enough for the Glauber-Gribov coupled-channel
multiple scattering theory (CCMST) \cite{Glauber,Gribov} to
be applicable for this purpose.
The CCMST allows one to sum the quantum mechanical amplitudes
contributing to electroexcitation
and diffractive de-excitation of the struck proton
$p\rightarrow i_{1}\rightarrow ...\rightarrow i_{\nu}\rightarrow p$
provided that the nucleus wave function, $3q$-nucleon scattering
matrix and initial ejectile $3q$ wave function
are known.
In CCMST the CT phenomenon corresponds to a cancellation
between the rescattering amplitudes with elastic (diagonal)
and inelastic (off-diagonal) intermediate states.
Such a nontrivial cancellation becomes possible in QCD due to
the existence of the CT sum rules \cite{OnsetCT}, which
relate diagonal and off-diagonal transition amplitudes.

Several works were devoted to the study of CT effects
in $(e,e'p)$ scattering within the coupled-channel formalism
under different assumptions for the $3q$-nucleon scattering
amplitudes \cite{OnsetCT,BBK1,JK,Jasym,BBK2}.
The results of these analyses show that in the case
of the integrated nuclear transparency effect of
the off-diagonal rescatterings is still small
at $Q^{2}\lsim 10$ GeV$^{2}$. More recent calculations
\cite{Kohama1,Kohama2}
within the Green's function approach developed in refs.
\cite{S3q1,S3q2} also yield
slow onset of CT. This prediction is consistent with
the weak $Q^{2}$-dependence of the nuclear transparency
observed in the NE18 experiment \cite{NE18}.

FSI effects vary with the missing momentum, $\vec{p}_{m}$,
and CT effects may be enhanced in some kinematical regions.
The analyses of refs. \cite{JK,Jasym,BBK2} show that the off-diagonal
rescatterings give rise to the forward-backward
(F-B) asymmetry of the missing momentum distribution.
In the forthcoming high precision experiments at CEBAF, such
a F-B asymmetry could have been a better signature of CT than
the weak $Q^{2}$-dependence of the integrated nuclear transparency.
However, the nonzero Re/Im ratio, $\alpha_{pN}$, for the forward
$pN$ scattering amplitude makes the nuclear medium dispersive
for the struck proton and leads
to a difference between its asymptotic momentum and momentum
inside the nucleus. This difference entails a longitudinal
shift of the missing momentum distribution and also generates
the F-B asymmetry which has not been considered in
\cite{JK,Jasym,BBK2}.
Within the context of the
analysis of the inclusive data from SLAC, the role
played by $\alpha_{pN}$ in $eA$ scattering at high $Q^{2}$ has been
pointed out in ref. \cite{gangofsix}.
The recent analyses \cite{NSZ,AZ} demonstrated that the F-B
asymmetry, associated with $\alpha_{pN}$,
at $Q^{2}
\mathrel{\rlap{\lower4pt\hbox{\hskip1pt$\sim$}} \raise1pt\hbox{$<$}}
10$ GeV$^{2}$ is of the same order, or even larger than
generated by the CT effects.
Besides omitting the large effect of $\alpha_{pN}$ on the F-B
asymmetry, CT effects in refs. \cite{JK,Jasym,BBK2}
were evaluated under certain qualitative approximations.
In ref. \cite{Jasym} the sum of the CCMST series was performed
using the approximation of the effective diffraction scattering
matrix; ref. \cite{JK} has used
several unjustified approximations in the numerical calculations, and
the authors of ref. \cite{BBK2} have used incorrect initial ejectile
$3q$ wave function (for the criticism to the
approaches of refs. \cite{JK,BBK2} see \cite{AZ}).
Furthermore, none of refs. \cite{JK,Jasym,BBK2} has discussed
the $p_{m\perp}$-dependence of
CT effects. On the whole, the theoretical
understanding of the $\vec{p}_{m}$-dependence of CT
effects is far from being complete and further
investigations of this problem are required.

In the present paper we study the missing momentum dependence of
the nuclear transparency
in the region of $p_{m}\lsim k_{F}$ (here $k_{F}\approx 250$ MeV/c is
the Fermi momentum).
We perform an exact evaluation of CCMST series, thus improving
upon the approximation of the effective diffraction matrix
\cite{Jasym}, and for the first time study the convergence of CCMST
expansion in the number of the excited proton states.
In our evaluation of CT effects we use
the realistic Pomeron part of the $3q$-nucleon diffraction
scattering matrix, which was previously
used in ref. \cite{OnsetCT} for calculation of the
integrated nuclear transparency. We study the sensitivity
of the results to the choice of the reggeon $3q$-nucleon amplitudes.

The present analysis is focused on the use of CCMST
to describe the evolution of the $3q$ ejectile state
during its propagation through the nuclear medium.
We evaluate CCMST series describing the nucleus wave function
in the independent particle shell model.
The short range correlations
(SRC) have been neglected
motivated by the relatively
small correlation effects on the SPMD \cite{NNcor,pieper}
and on the missing
momentum distribution in $^{4}He(p,2p)$
found in the recent
many-body Glauber analysis \cite{He42}
at $p_{m}\lsim k_{F}$.
However in ref. \cite{pieper}, it is observed that NN short range 
central correlation is responsible for the SPMD tail and that the tensor
correlation enhances the tail 
further almost by a factor of 3 without changing its shape.
It is feasible that using Monte Carlo approaches
for light nuclei and the local density approximation
for heavier systems, one could eventually perform a more
sophisticated analysis including SRC.

The coupled channel formalism is developed in a form
which includes both the coherent and incoherent rescatterings
of the ejectile state in the nuclear medium.
In the single-channel Glauber model, the role of the incoherent FSI
was elucidated in ref. \cite{NSZ}.
It was shown that in the shell model, the allowance for
both the coherent and incoherent rescatterings
corresponds to the inclusive $(e,e'p)$ reaction, when all the final
states of the residual nucleus are involved, while
the cross section obtained neglecting the incoherent
rescatterings is related to the
exclusive $(e,e'p)$ reaction, when only the one-hole
excitations of the target nucleus are allowed.
The analysis \cite{NSZ} shows that the incoherent
rescatterings become important
at $p_{m}\gsim 200-250$ MeV/c.
The impact of the incoherent off-diagonal rescatterings on
the onset of CT has not yet been treated quantitatively.
We demonstrate that, on the contrary to
the coherent off-diagonal rescatterings, the contribution of the
off-diagonal incoherent rescatterings does not vanish at low $Q^{2}$.
We also show that in the case of incoherent rescatterings
the CT effects decrease the nuclear transparency.

The numerical calculations of the present paper are performed
for the exclusive $(e,e'p)$ reaction.
In the region of the relatively small missing momenta
($p_{m}\lsim 150-200$ MeV/c), where the contribution of the incoherent
rescatterings is negligible, our theoretical
predictions may be compared with inclusive experimental data.

The correspondence between the coherent FSI and the one-hole
excitations
allows an evaluation of the $\vec{p}_{m}$-dependence of CT effects
for different hole states. Because of the change of the spatial
distribution of the bound proton, we find significant variations
of nuclear transparency from the one hole state to another.
We present estimates of CT effects for different acceptance
windows in the transverse and longitudinal missing momentum.
Previously, different hole excitations were considered in
\cite{Strikman} in a model of classical evolution of the ejectile state,
which conflicts the coherency properties of CCMST.

The paper is organized as follows.
In section 2 we set out the CCMST formalism
for $(e,e'p)$ reaction. The emphasis is
placed on the approximations which are necessary to obtain
the intuitive formula of the optical approximation.
The considerations of the parameterization of
the diffraction matrix and of the initial ejectile wave
function are given in section 3.
In section 4 we apply the formalism of CCMST for qualitative
analysis of the incoherent FSI.
The numerical results obtained for exclusive
$^{16}O(e,e'p)$ and $^{40}Ca(e,e'p)$ reactions
are presented in section 5.
The summary and conclusions are presented in section 6.

\section{FSI in the Glauber-Gribov formalism}
We begin with the kinematics of quasielastic $(e,e'p)$ scattering.
In the present paper we, following the usual practice
\cite{Forest,EEPR1,EEPR2}, assume that
at high $Q^{2}$ the differential cross section of $(e,e'p)$ reaction
may be expressed
through the half off-shell $ep$-cross section, $\sigma_{ep}$,
and the distorted spectral function, $S(E_{m},\vec{p}_{m})$,
as
\beq
{d\sigma \over dQ^{2}d\nu dp d\Omega_{p}}=
K
\sigma_{ep}
S(E_{m},\vec{p}_{m})\, .
\label{eq:1}
\endeq
Here $K$ is a kinematical factor,
$\nu$ and $\vec{q}$ are the $(e,e')$ energy
and momentum transfer, $Q^{2}=\vec{q}\, ^{2}-\nu^{2}$, the struck
proton has a momentum $\vec{p}$ and energy $E(p)=T_{kin}+m_{p}$,
the missing momentum and energy are defined as
$\vec{p}_{m}=
\vec{q}-\vec{p}
$ and $E_{m}=\nu+m_{p}-E(p)$ and
the $z$-axis is chosen along $\vec{q}$.
Apart from $E_{m}$ and $\vec{p}_{m}$ the distorted spectral function
depends on $\vec{p}$. In Eq.~(\ref{eq:1}) and hereafter we
suppress this variable.
Eq. (\ref{eq:1}) is written under the assumption that the difference
between the spectral functions corresponding to absorption of
the longitudinal (L) and transverse (T) photons, connected with
the spin dependence of FSI and CT effects,
can be neglected. We ignore the spin effects in FSI because
at large energy of the struck proton they become small.
As far as the CT effects are concerned, we will see that
in the case of dominance of the small-size $3q$ configurations
in hard $ep$ scattering, which is of our interest in the
present paper, the contribution of the off-diagonal rescatterings
to the longitudinal and transverse spectral functions
must be close to each other.
Since we do not distinguish the longitudinal and transverse
spectral functions, below we treat the
electromagnetic current as a scalar operator.
Also, notice that Eq.~(\ref{eq:1}) is for the cross section
averaged over the azimuthal angle between the missing
momentum and the $(e,e')$ reaction plane,
which does not contain the LT and TT interference responses
\cite{EEPR2}.

In terms of the distorted spectral function the nuclear transparency
for a certain kinematical domain, $D$, of the missing energy and
the missing momentum can be written as
\beq
T_{A}(D)=\frac{\int\limits_{D} dE_{m}d^{3}\vec{p}_{m}
S(E_{m},\vec{p}_{m})}
{\int\limits_{D} dE_{m}d^{3}\vec{p}_{m}
S_{PWIA}(E_{m},\vec{p}_{m})}\,.
\label{eq:2}
\eeq
Here $S_{PWIA}(E_{m},\vec{p}_{m})$ is the theoretical
spectral function of PWIA calculated without taking into account FSI.
The missing momentum distribution which is of our interest
in the present paper is given by
\beq
w(\vec{p}_{m})={1 \over (2\pi)^{3}} \int  dE_{m} S(E_{m},\vec{p}_{m})\,.
\label{eq:3}
\eeq

The distorted spectral function can be written as
\beq
S(E_{m},\vec{p}_{m})=\sum_{f}|M_{f}(\vec{p}_{m})|^{2}
\delta(E_{m}+E_{A-1}(\vec{p}_{m})+m_{p}-m_{A})\,,
\label{eq:4}
\eeq
where
$M_{f}(\vec{p}_{m})$ is the reduced matrix element of the
exclusive process
 $e\,+\,A_{i}\,\rightarrow\,\,e'\,+\,(A-1)_{f}\,+\,p$.
Then, the missing momentum distribution reads
\beq
w(\vec{p}_{m})=
{1\over (2\pi)^{3}}
\sum_{f}|M_{f}(\vec{p}_{m})|^{2}\,.
\label{eq:5}
\eeq

In our analysis we confine ourselves to a large mass
number of the target nucleus $A\gg 1$. Then, neglecting
the center of mass correlations we can write
$M_{f}(\vec{p}_{m})$ as
\beq
M_{f}(\vec{p}_{m})=\int d^{3}\vec{r}_{1}...d^{3}\vec{r}_{A}
\Psi_{f}^{*}(\vec{r}_{2},...,\vec{r}_{A})
\Psi_{i}(\vec{r}_{1},...,\vec{r}_{A})
S(\vec{r}_{1},...\vec{r}_{A})\exp(i\vec{p}_{m}\vec{r}_{1})\,.
\label{eq:6}
\eeq
Here $\Psi_{i}$ and $\Psi_{f}$ are wave functions
of the target and residual nucleus, respectively. The nucleon "1" is
chosen to be the struck proton.
For the sake of brevity, in Eq (\ref{eq:6}) and hereafter
the spin and isospin variables are suppressed.
The factor $S(\vec{r}_{1},...,\vec{r}_{A})$, which takes into
account FSI of the ejectile state with spectator nucleons,
is given by
\beq
S(\vec{r}_{1},\vec{r}_{2},...,\vec{r}_{A})=
\frac{\langle p|\hat{S}_{3q}(\vec{r}_{1},\vec{r}_{2},...,\vec{r}_{A})
|E\rangle}
{\langle p|E\rangle}\,,
\label{eq:7}
\eeq
where $|E\rangle$ is a three-quark wave function which
describes the state of
the proton after absorption of the virtual photon at point
$\vec{r}_{1}$ and
$\hat{S}_{3q}(\vec{r}_{1},\vec{r}_{2},...,\vec{r}_{A})$ is an
evolution operator of the three-quark system in the nuclear medium
(as usual we assume that the spectator coordinates
may be considered frozen during propagation of the fast
$3q$ system through the nuclear medium).
In the right-hand side of Eq.~(\ref{eq:7}) the numerator is
the probability
amplitude for the $3q$ ejectile state escaping from
the target nucleus $(A-1)_{f}$ debris
to be observed in the proton state $|p\rangle$, and
the denominator is the probability amplitude
for the state $|E\rangle$ to be observed in the proton state
as well.
In terms of the electromagnetic current operator
$\hat J_{em}$, the ejectile wave function is expressed as
\cite{Jasym}
\beq
|E\rangle=\hat{J}_{em}(Q)|p\rangle=\sum\limits_{i}
|i\rangle\langle i|J_{em}(Q)|p\rangle=\sum\limits_{i}
G_{ip}(Q)|i\rangle\,,
\label{eq:8}
\eeq
where $G_{ip}(Q)=\langle i|J_{em}(Q)|p\rangle$ includes the
electromagnetic form factor of the proton as well as all
the transition form factors for the electroexcitation of the proton
$e~+~p~\rightarrow~e'~+~i$.
It is worth noting that the ejectile wave function (\ref{eq:8})
is independent of
the missing momentum and $\vec{p}_{m}$-dependence of the
reduced matrix element emerges only through the
exponential $\exp(i\vec{p}_{m}\vec{r}_{1})$ in the right-hand side of
Eq.~(\ref{eq:6}).

In the coupled-channel formalism the evolution operator of
the $3q$ system can be written in the following form
\beq
\hat{S}_{3q}(\vec{r}_{1},...,\vec{r}_{A})=
\hat{P}_{z}\prod\limits_{j=2}^{A}
\left[1-\theta(z_{j}-z_{1})
\hat{\Gamma}(\vec{b}_{j}-\vec{b}_{1},z_{j}-z_{1})\right]\,,
\label{eq:9}
\eeq
where $\vec{b}_{j}$ and $z_{j}$ are the transverse and
longitudinal coordinates of the nucleons,
$\hat\Gamma(\vec{b},z)$ is the operator profile function
describing $3q$-nucleon scattering.
At high energy the $3q$ system
propagates along the straight-path trajectory and
can interact with the spectator nucleon "j"
only provided that $z_{j}>z_{1}$, which is an origin of
the $z$-ordering operator $\hat{P}_{z}$ and of the step-function
$\theta(z_{j}-z_{1})$ in Eq. (\ref{eq:9}).
The matrix elements of the $z$-dependent operator profile function
$\hat\Gamma(\vec{b},z)$ can be written as
\beq
\langle i|\hat\Gamma(\vec{b},z)|j\rangle
=\exp(ik_{ij}z)\langle i|\hat\Gamma(\vec{b})|j\rangle\,,
\label{eq:10}
\eeq
where  $\hat\Gamma(\vec{b})$ is the usual operator profile
function connected with the scattering matrix, $\hat f$,
through the relation
\beq
\langle i|\hat\Gamma(\vec{b})|j\rangle=
-\frac{i}{8\pi^{2}}
\int d^{2}\vec{q}\exp(i\vec{q}\,\vec{b})
\langle i|\hat f(\vec{q}\,)|j\rangle\,
\label{eq:11}
\eeq
(the normalization of the scattering matrix is such
that Im$\langle i|\hat{f}(\vec{q}=0)|i\rangle=\sigma_{tot}(iN)$)
, and
$k_{ij}$ is the longitudinal momentum transfer
related to transition $i N\rightarrow j N$ \cite{Gribov}
\beq
k_{ij}=\frac{m_{i}^{2}-m_{j}^{2}}{2\varepsilon}\,,
\label{eq:12}
\eeq
here $\varepsilon$ is the energy of the struck proton in
the laboratory frame, $m_{i}$ and $m_{j}$ are the masses
of the states $|i\rangle$ and $|j\rangle$ . The exponential phase
factor in Eq.~(\ref{eq:10}) results from
the additional phase which the $3q$ plane wave
acquires after propagating the distance $z$.
The whole phase factor, which the operator (\ref{eq:9}) yields
in the case of
the sequence of intermediate states
$i_{1}\rightarrow ...\rightarrow i_{\nu}\rightarrow p$,
is given by
\beq
F(i_{1}\rightarrow ...\rightarrow i_{\nu}\rightarrow p)=
\exp\left[i\sum\limits_{j=1}^{\nu}
k_{p i_{j}}(z_{j}-z_{j-1})\right]\,,
\label{eq:13}
\eeq
where $z_{j}$ ($j\ge 1$) is the longitudinal coordinate of the point
where the transition $i_{j} N\rightarrow i_{j+1} N$
takes place, and $z_{0}$ corresponds to
the transition $p+\gamma^{*}\rightarrow i_{1}$.
It is easy to check, that the same
phase factor (\ref{eq:13}) can be obtained by solving the set of
the coupled-channel wave equations.

The sum over the final states of the residual nucleus in Eq.~(\ref{eq:5})
can be performed with the help of the closure relation
\beq
\sum_{f}\Psi_{f}(\vec{r}_{2}^{\,'},...,\vec{r}_{A}^{\,'})
\Psi^{*}_{f}(\vec{r}_{2},...,\vec{r}_{A})=
\prod\limits_{j=2}^{A}\delta(\vec{r}_{j}-\vec{r}_{j}^{\,'})  \,.
\label{eq:14}
\eeq
After
making use of (\ref{eq:6}), (\ref{eq:14}),
the missing momentum distribution (\ref{eq:5}) can be cast in the
form
\beq
w(\vec{p}_{m})=\frac{1}{(2\pi)^{3}}\int
d^{3}\vec{r}_{1}d^{3}\vec{r}_{1}^{\,'}
\rho_{D}(\vec{r}_{1},\vec{r}_{1}^{\,'})
\exp[i\vec{p}_{m}(\vec{r}_{1}-\vec{r}_{1}^{\,'})]\,,
\label{eq:15}
\eeq
where
\beq
\rho_{D}(\vec{r}_{1},\vec{r}_{1}^{\,'})=
\int
\prod\limits_{j=2}^{A}d^{3}\vec{r}_{j}
\Psi_{i}(\vec{r}_{1},\vec{r}_{2},...,\vec{r}_{A})
\Psi_{i}^{*}(\vec{r}_{1}^{\,'},\vec{r}_{2},...,\vec{r}_{A})
S(\vec{r}_{1},\vec{r}_{2},...,\vec{r}_{A})
S^{*}(\vec{r}_{1}^{\,'},\vec{r}_{2},...,\vec{r}_{A})\,.
\label{eq:16}
\eeq
 The function $\rho_{D}(\vec{r}_{1},\vec{r}_{1}^{\,'})$
can be viewed as a FSI-modified one-body proton density matrix.
In PWIA, when  the FSI factors in the right-hand side
of Eq.~(\ref{eq:16}) equal unity, (\ref{eq:16})
reduces to the formula	for usual one-body
proton density matrix $\rho(\vec{r}_{1},\vec{r}_{1}^{\,'})$,
and Eq.~(\ref{eq:15}) reduces to the expression for SPMD
\beq
n_{F}(\vec{p}_{m})
=
{1\over (2\pi)^{3}}\int
 d\vec{r}_{1}d\vec{r}_{1}^{\,'}
 \rho(\vec{r}_{1},\vec{r}_{1}^{\,'})
\exp\left[i\vec{p}_{m}(\vec{r}_{1}-\vec{r}_{1}^{\,'})\right] \,.
\label{eq:17}
\eeq

As was stated in section 1, we will describe the target
nucleus in the independent particle shell model.
After neglecting the SRC
the $A$-body semidiagonal density matrix
$\Psi_{i}(\vec{r}_{1},\vec{r}_{2},...,\vec{r}_{A})
\Psi_{i}^{*}(\vec{r}_{1}^{\,'},\vec{r}_{2},...,\vec{r}_{A})$
in Eq.~(\ref{eq:16})
still contains the Fermi correlations.
To carry out the integration over the coordinates of the
spectator nucleons we neglect the Fermi correlations and
replace the $A$-body semidiagonal density matrix by the
factorized form
\beq
\Psi_{i}(\vec{r}_{1},\vec{r}_{2},...,\vec{r}_{A})
\Psi_{i}^{*}(\vec{r}_{1}^{\,'},\vec{r}_{2},...,\vec{r}_{A})
\rightarrow
\rho(\vec{r}_{1},\vec{r}_{1}^{\,'})\prod\limits_{i=2}^{A}
\rho(\vec{r}_{i})
\,.
\label{eq:18}
\eeq
Here
$$
\rho(\vec{r}_{1}, \vec{r}_{1}^{\,'})=\frac{1}{Z} \sum_{n}
\phi_{n}^{*}(\vec{r}_{1}^{\,'})\phi_{n} ( \vec{r}_{1})
$$
is the proton shell model one-body density matrix and
$\phi_{n}$ are the shell model wave functions,
$\rho_{A}(\vec{r}\,)$ is the normalized to unity nucleon nuclear
density. The errors connected with ignoring the Fermi correlations
must be small because
the ratio between the Fermi correlation length $l_{F}\sim 3/k_{F}$
and the interaction length corresponding to the interaction of the
struck proton with the Fermi correlated spectator nucleons
$l_{int}\sim 4(\sigma_{tot}(pN)\langle n_{A}\rangle)^{-1}$
(here $\langle n_{A}\rangle$ is the average nucleon nuclear density)
is a small quantity ($\sim 0.25$).
Recall that
the factored approximation for the many-body
nuclear density has been successfully employed,
in connection with Glauber theory, in the analysis
of a wealth of hadron-nucleus scattering data
(for an extensive review on $hA$
scattering see \cite{Alkhaz}).

Making use of replacement (\ref{eq:18}) in Eq.~(\ref{eq:16})
allows to write the missing momentum
distribution in the form
\beq
w(\vec{p}_{m})=\frac{1}{(2\pi)^{3}}\int
d^{3}\vec{r}_{1}d^{3}\vec{r}_{1}^{\,'}
\rho(\vec{r}_{1},\vec{r}_{1}^{\,'})
\Phi(\vec{r}_{1},\vec{r}_{1}^{\,'})
\exp[i\vec{p}_{m}(\vec{r}_{1}-\vec{r}_{1}^{\,'})]\,,
\label{eq:20}
\eeq
where
the FSI factor $\Phi(\vec{r}_{1},\vec{r}_{1}^{\,'})$
is given by
\beq
\Phi(\vec{r}_{1},\vec{r}_{1}^{\,'})=
\int \prod\limits_{j=2}^{A}\rho_{A}(\vec{r}_{j}) d^{3}\vec{r}_{j}
S(\vec{r}_{1},\vec{r}_{2},...,\vec{r}_{A})
S^{*}(\vec{r}_{1}^{\,'},\vec{r}_{2},...,\vec{r}_{A})\,.
\label{eq:21}
\eeq

Substituting (\ref{eq:7}) into (\ref{eq:21}) we may write the FSI factor
$\Phi(\vec{r}_{1},\vec{r}_{1}^{\,'})$  in the following
operator form
\beq
\Phi(\vec{r}_{1},\vec{r}_{1}^{\,'})=
\langle p p|\hat{U}(\vec{r}_{1},\vec{r}_{1}^{\,'})|
i i^{'}\rangle C_{i}C_{i'}^{*}\,\,,
\label{eq:22}
\eeq
where
$$
C_{i}=\frac{\langle i|E\rangle}{\langle p|E\rangle}\,,
$$
and the operator $\hat{U}(\vec{r}_{1},\vec{r}_{1}^{\,'})$ is an evolution
operator for the density matrix
of the $3q$ system in the nuclear medium
\beq
\hat{U}(\vec{r}_{1},\vec{r}_{1}^{\,'})=
\int \prod\limits_{j=2}^{A}\rho_{A}(\vec{r}_{j}) d^{3}\vec{r}_{j}
\hat{S}_{3q}(\vec{r}_{1},\vec{r}_{2},...,\vec{r}_{A})
\hat{S}_{3q}^{*}(\vec{r}_{1}^{\,'},\vec{r}_{2},...,\vec{r}_{A})\,.
\label{eq:23}
\eeq
After substitution
of the CCMST expression (\ref{eq:9}) for $\hat{S}_{3q}$ into
Eq.~(\ref{eq:23}) it takes the form
\bea
\hat{U}(\vec{r}_{1},\vec{r}_{1}^{\,'})=\hat{P}_{z}\left[1-
\int\limits_{z_{1}}^{\infty}d z
\int
d^{2}\vec{b}\,\hat{\Gamma}(\vec{b}-\vec{b}_{1},z-z_{1})
\rho_{A}(\vec{b},z)
\right.\nonumber\;\;\;\;\;\;\;\;\;\;\;\;\;\;\;\;\;\;\;\;\\ \left.
-\int\limits_{z_{1}^{'}}^{\infty}d z
\int
d^{2}\vec{b}\,\hat{\Gamma}^{*}(\vec{b}-\vec{b}_{1}^{\,'},z-z_{1}^{'})
\rho_{A}(\vec{b},z)
\right.\nonumber\;\;\;\;\;\;\;\;\;\;\;\;\;\;
\\ \left. +\int\limits_{max(z_{1},z_{1}^{\,'})}^{\infty}
d z\int d^{2}\vec{b}\,
\hat{\Gamma}(\vec{b}-\vec{b}_{1},z-z_{1})
\hat{\Gamma}^{*}(\vec{b}-\vec{b}_{1}^{\,'},z-z_{1}^{'})
\rho_{A}(\vec{b},z)\right]^{A-1}.
\label{eq:24}
\eea
We will refer to the first two terms in the square brackets
in the right-hand side of (\ref{eq:24}) as \G terms, and to the last one
as \GG term. The operator (\ref{eq:24}) can be graphically represented by
the sum of the diagrams as shown in Fig.~1.
Every dotted line attached to the straight-line trajectory originating from
 the point $\vec{r}_{1}$ ($\vec{r}_{1}^{\,'}$)
 denotes a profile function
$\hat\Gamma(\vec{b}_{j}-\vec{b}_{1},z_{j}-z_{1})$
($\hat\Gamma^{*}(\vec{b}_{j}-\vec{b}_{1}^{\,'},z_{j}-z_{1}^{'})$).
The interaction between the two trajectories
generated by the diagrams like shown in
Fig.~1b
does not allow one to represent (\ref{eq:24}) in a factorized
form in coordinate and internal space of the two $3q$ systems propagating
along the trajectories originating from $\vec{r}_{1}$ and
$\vec{r}_{1}^{\,'}$.

The emergence of the evolution operator for the $3q$ density
matrix in Eq.~(\ref{eq:22}) is not surprising
since we evaluate the probability distribution for a subsystem
(the $3q$ ejectile state) in the process involving a complex system
(the $3q$ ejectile state and the residual nucleus). It is
worth recalling that
similar interaction between the trajectories
emerges also in the related problem of passage of the
ultrarelativistic positronium through matter
\cite{A2e}.

Eq.~(\ref{eq:24}) can further be simplified
exploiting the fact
that $\rho_{A}(\vec{b},z)$ is a smooth function of the impact
parameter $\vec{b}$ as compared to the operator profile
function. In leading order in the small parameter
$R_{3qN}^{2}/R_{A}^{2}$
($R_{3qN}$ is a radius of  $3q$-nucleon interaction  and $R_{A}$ is
 the nucleus radius) we have
\beq
\int d^{2}\vec{b}\,\hat{\Gamma}(\vec{b}-\vec{b}_{1},z-z_{1})
\rho_{A}(\vec{b},z)\approx
\frac{\hat{\sigma}(z)}{2}\rho_{A}(\vec{b},z)\,,
\label{eq:25}
\eeq
\bea
\int d^{2}\vec{b}\,
\hat{\Gamma}(\vec{b}-\vec{b}_{1},z-z_{1})
\hat{\Gamma}^{*}(\vec{b}-\vec{b}_{1}^{\,'},z-z_{1}^{'})
\rho_{A}(\vec{b},z)
\;\;\;\;\;\;\;\;\;\;\;\;\;\;\;\;\;\;\;\;\;\;\;\;\;\nonumber\\
\approx
\hat{\eta}(\vec{b}_{1}-\vec{b}_{1}^{\,'},z-z_{1},z-z_{1}^{'})
\rho_{A}(\frac{1}{2}(\vec{b}_{1}+\vec{b}_{1}^{\,'}),z)\,\,,
\label{eq:26}
\eea
where
\beq
\hat{\sigma}(z)=2\int d^{2}\vec{b}\,\hat{\Gamma}(\vec{b},z)\,,
\label{eq:27}
\eeq
\beq
\hat{\eta}(\vec{b},z,z')=\int
d^{2}\vec{\Delta}\hat{\Gamma}(\vec{b}-\vec{\Delta},z)
\hat{\Gamma}^{*}(\vec{\Delta},z')\,.
\label{eq:28}
\eeq
In terms of the diffraction scattering matrix $\hat{f}$
the matrix elements of the $z$-dependent operators (\ref{eq:27}),
(\ref{eq:28}) are given by
\beq
\langle i|\hat{\sigma}(z)|k\rangle=-i
\exp(ik_{ij}z)\langle i|\hat{f}(\vec{q}=0)|k\rangle\,,
\label{eq:29}
\eeq
\beq
\langle ii'|\hat{\eta}(\vec{b},z,z')|kk'\rangle=
\frac{\exp[i(k_{ik}z-k_{i'k'}z')]}{16\pi^{2}}
\int
d^{2}\vec{q}\exp(i\vec{q}\,\vec{b})
\langle i|\hat{f}(\vec{q}\,)|k\rangle
\langle i'|\hat{f}(-\vec{q}\,)|k'\rangle^{*}\,.
\label{eq:30}
\eeq
The results of the analysis of $(e,e'p)$ scattering within
the Glauber model \cite{NSZ} indicate that the variations of the
missing momentum distribution
connected with the smearing corrections to the approximations
(\ref{eq:25}), (\ref{eq:26}) are $\lsim 3$\% at $p_{m}\lsim 300$
MeV/c.

Making use of Eqs.~(\ref{eq:24})-(\ref{eq:26}) and exponentiating which
is a good approximation at $A\gsim 10$
we finally get
\bea
\hat{U}(\vec{r}_{1},\vec{r}_{1}^{\,'})=\hat{P}_{z}\exp\left[-\frac{1}{
2 } \int\limits_{z_{1}}^{\infty}d z\,
\hat{\sigma}(z-z_{1})
n_{A}(\vec{b}_{1},z)
-\frac{1}{2}\int\limits_{z_{1}^{'}}^{\infty}d z\,
\hat{\sigma}^{*}(z-z_{1}^{'})
n_{A}(\vec{b}_{1}^{\,'},z)
\right.\nonumber\\ \left. +\int\limits_{max(z_{1},z_{1}^{\,'})}^{\infty}
dz\,\hat{\eta}(\vec{b}_{1}-\vec{b}_{1}^{\,'},z-z_{1},z-z_{1}')
n_{A}(\frac{1}{2}(\vec{b}_{1}+\vec{b}_{1}^{\,'}),z)\right]\,,
\;\;\;\;\;\;\;
\label{eq:31}
\eea
where $n_{A}(\vec{r}\,)=A\rho_{A}(\vec{r}\,)$ is the nucleon nuclear
density.

As one can see from Eq.~(\ref{eq:28}) the \GG term in Eqs. (\ref{eq:24}),
(\ref{eq:31})
becomes only important when
$|\vec{b}_{1}-\vec{b}_{1}^{\,'}|\lsim R_{3qN}\sim 1$ fm.
Such a short range interaction between the two trajectories
in the impact parameter plane must for the most part
affect the missing momentum distribution at
$p_{m}\gsim 1/R_{3qN}\sim 200$ MeV/c.
This fact becomes evident if one rewrites Eq.~(\ref{eq:20}) in the
convolution form
\beq
w(\vec{p}_{m})=
\frac{1}{(2\pi)^{6}}
\int
d^{3}\vec{R}
\int d^{3}\vec{k}
W_{\rho}(\vec{R},\vec{p}_{m}-\vec{k})W_{\Phi}(\vec{R},\vec{k})\,,
\label{eq:32}
\eeq
where
\beq
W_{\rho}(\vec{R},\vec{k})=\int
d^{3}\vec{r}\rho(\vec{R}+\vec{r}/2,\vec{R}-\vec{r}/2)
\exp(i\vec{k}\vec{r}\,)
\label{eq:33}
\eeq
is the familiar Wigner function, and
\beq
W_{\Phi}(\vec{R},\vec{k})=
\int
d^{3}\vec{r}\Phi(\vec{R}+\vec{r}/2,\vec{R}-\vec{r}/2)
\exp(i\vec{k}\vec{r}\,)\,.
\label{eq:34}
\eeq
The representation (\ref{eq:32}) makes it clear that the
short range interaction between
the two trajectories generated by the \GG term, which converts
after Fourier transform
(\ref{eq:34}) into the slowly decreasing tails of
$W_{\Phi}(\vec{R},\vec{k})$,
will reveal itself for the most part at large missing momenta.
Remarkably, although the
operator $\hat\eta(\vec{b},z,z')$ defined by Eq.~(\ref{eq:28}) has
the short range behavior only in the impact parameter plane,
the \GG term in Eq.~(\ref{eq:31}) generates the slow decreasing tails
of $W_{\Phi}(\vec{R},\vec{k})$ in the longitudinal momenta
as well. A formal origin of this effect is the
non-analytical behavior at $z_{1}=z^{'}_{1}$ of the function
$max(z_{1},z_{1}')=(z_{1}+z^{'}_{1}+|z_{1}-z^{'}_{1}|)/2$,
which is the low limit of
integration over $z$ in the \GG term in Eq.~(\ref{eq:31}).
Such a non-analytical function derives from the absence
of an incoming proton plane wave,
which is a real
physical reason for affecting the longitudinal missing momentum
distribution by the \GG term (for the detailed
quantum mechanical analysis of this phenomenon
see ref. \cite{NSZ}).

In the Glauber analysis of $(e,e'p)$ scattering \cite{NSZ} it was
shown that
the \GG-generated effects correspond to  the incoherent
rescatterings
of the struck proton in the nuclear medium, while the \G terms
describe FSI related to the coherent rescatterings.
>From the point of view of the shell model the theoretical predictions
obtained without taking into account the \GG term correspond to
the exclusive $(e,e'p)$ scattering, when only
the one-hole excitations of the target nucleus are allowed
, while the whole FSI factor including
the \GG term corresponds to the inclusive process, when all the
final states of the residual nucleus are included \cite{NSZ}.
The results of ref. \cite{NSZ} show that the \GG term
increases the
missing momentum distribution by 3-7\% at $|\vec{p}_{m}|\lsim 250$
MeV/c in the parallel kinematics. For
the transverse
kinematics the same estimate is valid for $p_{m\perp}\lsim 200$ MeV/c.
The \GG term becomes especially important in the
region of $p_{m\perp}\gsim 250$ MeV/c, where it dominates in the
missing momentum distribution.
Evidently, approximately the same situation will take place in
CCMST in the regime of the onset of CT.

The theoretical study of CT effects including the \GG term
in the region of large transverse missing momenta
would be of great interest because in this case the missing
momentum distribution probes
the $3q$-nucleon scattering amplitude when the $3q$ wave function
is still close to the initial ejectile wave function.
Unfortunately, the calculations including the \GG term
require the information about $3q$-nucleon diffraction scattering
matrix at arbitrary momentum transfer, as one can see from
Eq.~(\ref{eq:30}). It renders difficult an accurate estimate
of the CT effects for the inclusive $(e,e'p)$
reaction in the region of large $p_{m}$.
Still, even a qualitative understanding of
the role of the \GG term
is interesting and we comment on that in
section 4.
We postpone a detail analysis of the inclusive reaction at large
$p_{m}$ for further publications.

In the present paper we focus on the numerical calculations
of the missing momentum distribution for exclusive
$(e,e'p)$ reaction, when the FSI is exhausted by
the coherent rescatterings. The corresponding
FSI factor (we label it as $\Phi_{coh}$), which may be obtained
from Eq.~(\ref{eq:21}) after neglecting the
\GG term in the evolution operator (\ref{eq:31}),
has the following
factorized form
\beq
\Phi_{coh}(\vec{r}_{1},\vec{r}_{1}^{\,'})=
S_{coh}(\vec{r}_{1})S_{coh}(\vec{r}_{1}^{\,'})^{*}\,
\label{eq:35}
\eeq
where
\beq
S_{coh}(\vec{r}_{1})=
\langle p|\hat{P}_{z}\exp\left[-\frac{1}{2}
\int\limits_{z_{1}}^{\infty}d z
\hat{\sigma}(z-z_{1})
n_{A}(\vec{b}_{1},z)\right]|i\rangle C_{i}\,\,.
\label{eq:36}
\eeq
Substituting (\ref{eq:35}) into (\ref{eq:20}) we arrive at the following
expression for the missing momentum distribution
\beq
w(\vec{p}_{m})=
\frac{1}{Z}\sum\limits_{n}
\left|\int
d^{3}\vec{r}\,\phi_{n}(\vec{r}\,)\exp(i\vec{p}_{m}\vec{r}\,)
S_{coh}(\vec{r}\,)\right|^{2}\,.
\label{eq:37}
\eeq
Eq.~(\ref{eq:37}) can be also obtained directly from
Eqs.~(\ref{eq:5}),~(\ref{eq:6})
if one includes in the sum over the final states of the residual
nucleus in Eq.~(\ref{eq:5}) only the one-hole
excitations and neglects the
Fermi correlations between the spectator nucleons.
For the related intuitive optical potential consideration
see refs. \cite{BBK1,JK,BBK2}.

>From the point of view of numerical calculations
it is convenient to evaluate $S_{coh}(\vec{r}\,)$
treating in Eq.~(\ref{eq:36})
the nondiagonal part of matrix
$\hat{\sigma}(z-z_{1})$ as a perturbation. Then, the FSI factor
(\ref{eq:36}) can be expanded
in the $\nu$-fold off-diagonal rescatterings series
\beq
S_{coh}(\vec{r}\,)=
\sum\limits_{\nu=0}^{\infty}
S_{coh}^{(\nu)}(\vec{r}\,)\,,
\label{eq:38}
\eeq
where
\beq
S_{coh}^{(0)}(\vec{r}\,)=\exp[-{1\over 2}t(\vec{b},\infty,z)
\sigma_{pp}]\,,
\label{eq:39}
\eeq
 and
\bea
S_{coh}^{(\nu)}(\vec{b},z)=\left(-\frac{1}{2}\right)^{\nu}
\sum\limits_{i_{1},...i_{\nu}}\sigma^{'}_{pi_{\nu}}
\sigma^{'}_{i_{\nu}i_{\nu-1}}\cdots
\sigma^{'}_{i_{2}i_{1}}
\frac{\langle i_{1}|E\rangle}{\langle p|E\rangle}
\exp[i k_{i_{1}p}z]
\int\limits_{z}^{\infty}dz_{1} n_{A}(\vec{b},z_{1})
\,\,\,\,\,\,\,\,\,\,\nonumber\\ \times
\exp[i k_{i_{2}i_{1}}z_{1}-{1\over 2}
t(\vec{b},z_{1},z)\sigma_{i_{1}i_{1}}]
\int\limits_{z_{1}}^{\infty}
dz_{2} n_{A}(\vec{b},z_{2})
\exp[i k_{i_{3}i_{2}}z_{2}-{1\over 2}
t(\vec{b},z_{2},z_{1})\sigma_{i_{2}i_{2}}]
\cdots
\,\,\nonumber\\
\times
\int\limits_{z_{\nu-1}}^{\infty}
dz_{\nu}
n_{A}(\vec{b},z_{\nu})
\exp[i k_{pi_{\nu}}z_{\nu}-{1\over 2}
t(\vec{b},\infty,z_{\nu})
\sigma_{pp}]\,, \;\;\;\,\,\,\,\,\,\,\,\nu\ge 1\,.
\label{eq:40}
\eea
Here $\vec{r}=(\vec{b},z)$,
$\sigma^{'}_{ik}=\sigma_{ik}-\delta_{ik}\sigma_{ii}$,
the matrix $\hat{\sigma}$
is connected with the forward diffraction
scattering matrix $\hat{f}(\vec{q}=0)=i\hat{\sigma}$
and
$t(\vec{b},z_{2},z_{1})=\int_{z_{1}}^{z_{2}}dz n_{A}(\vec{b},z)\,$
is the partial optical thickness.
The zeroth order term
$S^{(0)}(\vec{r}\,)$ in Eq.~(\ref{eq:38}) describes
the conventional Glauber
result,  while the terms with $\nu\ge 1$ correspond to the
inelastic intermediate states contributing to
electroexcitation and diffractive de-excitation
of the proton $p\rightarrow i_{1}\rightarrow ...\rightarrow
i_{\nu} \rightarrow p$.
It is precisely the oscillating exponential phase factors
in Eq.~(\ref{eq:40}), which
leads to suppression of the contributions of the inelastic
intermediate states at low energies of the struck proton.
They are also the origin of the longitudinal asymmetry of the
nuclear transparency produced by the off-diagonal rescatterings.
The emergence of these oscillating
factors is a purely quantum mechanical
effect. In the classical treatment of FSI in terms of
$z$-dependent $3q$-nucleon cross section of ref. \cite{Strikman}
it is lacking and the longitudinal asymmetry
of the missing momentum distribution vanishes.

Eqs.~(\ref{eq:37})-(\ref{eq:40}) form a basis for evaluation of the missing
momentum distribution within CCMST
for the exclusive $(e,e'p)$ reaction.
In the region of $p_{m}\lsim 150-200$ MeV/c, where the effect
of the incoherent rescatterings becomes small,
our predictions can be compared directly with experimental data
obtained without restrictions
on the final states of the residual nucleus.
\section{Parameterization of the diffraction matrix
 and the initial ejectile wave function}
To proceed further with the numerical calculation of
nuclear transparency we need the
diffraction matrix describing $3q$-nucleon scattering and
initial ejectile wave function.
At GeV's energies of the
struck proton, which are of our interest in the present paper,
the major contribution to the imaginary part of the amplitude
$f(kN\rightarrow iN)$ comes from the Pomeron exchange.
Namely this component of the $3q$-nucleon
scattering matrix
is for the most part important from the point of view of CT.
Following ref. \cite{OnsetCT} we
construct the Pomeron component
of $\hat\sigma$ using the oscillator quark-diquark model of the proton:
\beq
{\rm Im}~f_{P}(kN\rightarrow iN)=
{\rm Re}~\sigma_{ik}^{P}=\int dz d^{2} \vec{\rho} \,
\Psi_{i}^{*}(\vec{\rho},z)\sigma(\rho)\Psi_{k}(\vec{\rho},z)\,,
\label{eq:41}
\eeq
where $\Psi_{i,k}(\vec{\rho},z)$ are the oscillator wave functions describing
the quark-diquark states and $\sigma(\rho)$ is the dipole cross
section describing the interaction of the quark-diquark system
with a nucleon.
For the oscillator frequency of the quark-diquark system we
use the value $\omega_{qD}=0.35$ GeV,
 leading to a realistic mass spectrum of the proton
excitations.
As in ref. \cite{OnsetCT}
we take the dipole cross section in the form
\beq
\sigma(\rho)=\sigma_{0}\left [
1-\exp\left (-\frac{\rho^{2}}{R_{0}^{2}}\right )\right ]\,.
\label{eq:42}
\eeq
Eq. (\ref{eq:42}) is motivated by the results of the calculation
of the $q\bar{q}$ dipole cross section in the double gluon
exchange model of the Pomeron \cite{Low}. The parameterization
(\ref{eq:42}) with $\sigma_{0}\approx 50-60$ mb and
$R_{0}\approx 1.2-1.4$ fm allows one to describe both the CT effects
in
quasielastic charge exchange reaction $\pi^{-}A\rightarrow \pi^{o}A'$
\cite{dipsig1} and the nuclear shadowing and diffraction cross section
in deep inelastic scattering \cite{dipsig2,dipsig3}.
The $q\bar{q}$ dipole cross section
extracted from the experimental data on the vector meson electroproduction
\cite{dipsig4} also appears to be close to the one used in refs.
\cite{dipsig1,dipsig2,dipsig3}.
Of course, due to the nonzero diquark size, the
parameters of the quark-diquark dipole cross section
in Eq.~(\ref{eq:42}) may differ from the ones obtained from
the analysis of meson exchange process and deep inelastic scattering.
Following ref. \cite{OnsetCT}, we set $\sigma_{0}=80$ mb
and adjust $R_{0}$ to reproduce $\sigma_{tot}^{exp}(pN)$.
>From the point of view of realistic evaluation of the
onset
of CT it is important for the model diffraction matrix to
reproduce the gross features of diffractive $pN$ scattering in the
resonance region.
Our diffraction matrix obtained with the above set
of parameters yields the value of the ratio between
the diffractive and elastic $pN$ cross sections
$\sigma_{diff}(pp)/\sigma_{el}(pp)
\approx 0.25$, which is in agreement with
the experimental data \cite{DD}.
Moreover, we obtain a good description of the diffractive mass
spectrum observed in $pN$ scattering \cite{NNZ}.

The real parts of the diagonal $f(i N\rightarrow i N)$, for
$i\ne p$, and off-diagonal $f(i N\rightarrow k N)$ amplitudes are
not known experimentally. At GeV's energy of the struck proton
they are connected with the reggeon exchanges.
In the counterdistinction to the
Pomeron exchange we do not have at present a reliable theoretical
model for the reggeon contribution to the
$3q$-nucleon amplitudes even for a small-size $3q$ system.
As a reference value, we consider the
choice $\alpha_{1}=1$, $\alpha_{2}=0$ in
the parameterizations
\bea
{\rm Re}~f_{R}(iN\rightarrow iN)=
\alpha_{1}{\rm Re}~f_{R}(pN\rightarrow pN)=
\frac{\alpha_{1}}{2}
\left(\alpha_{pp}\sigma_{tot}(pp)+
\alpha_{pn}\sigma_{tot}(pn)\right)\,,\nonumber\\
{\rm Re}~f_{R}(iN\rightarrow k N)=
\alpha_{2}{\rm Im}~f_{P}(iN\rightarrow k
N)\,,\;\;\;\;i\neq k\,.\;\;\;\;\;\;\;\;\;\;\;
\label{reggeon}
\eea
Even though the above choice of $\alpha_{1,2}$
can be justified within the framework of the dual
parton model \cite{DTU} we are fully aware that it should only be
regarded as an estimate, and will study the
sensitivity of the results to the values of
$\alpha_{1,2}$.

Besides the matrix $\hat\sigma$, the evaluation of
$w(\vec{p}_{m})$ requires the initial ejectile wave function $|E\rangle$.
In the present paper we optimize for CT effects,
assuming the dominance of the small-size $3q$ configurations in the
matrix elements
$\langle i|J_{em}(Q)|p\rangle$ for the resonance states with
masses in the GeV's region \cite{Brodsky1}.
This amounts to a strong assumption that
the probability amplitude for the initial ejectile state
to be observed in state $|i\rangle$,
\beq
\langle i|E\rangle=\langle i|J_{em}(Q)|p\rangle
\propto \phi_{i}^{*}(\rho\sim \frac{1}{Q})\,,
\label{eq:43}
\eeq
where $\phi_{i}$ is the coordinate wave function of the state $|i\rangle$.
By virtue of Eq.~(\ref{eq:43})
the initial ejectile wave function can be chosen in a point-like
form. We parametrize it in the form
$\langle\rho|E\rangle\propto \exp(-C\rho^{2}Q^{2})$,
with $C=1$.
We would like to emphasize that the possibility of using
the point-like initial ejectile wave function
for the study of CT effects by no means
implies that the real ejectile state $|E\rangle$ actually
has a size $\sim 1/Q$. On the contrary, it is evident that
the ejectile state formed after absorption of the virtual
photon has exactly the same transverse size as the proton \cite{NNNJETP}.
The solution to this puzzling situation is
obvious.
Eq.~(\ref{eq:43}) is only valid for electroproduction of the proton and its
low-mass excitations,
which requires the hard gluon exchanges between the quarks of the $3q$ system.
The electroproduction of the high mass states with
masses $\sim |\vec{q}\,|$ does not require such exchanges, and
Eq.~(\ref{eq:43}) does not hold in this case.
However, the off-diagonal rescatterings including
the heavy intermediate states are suppressed.
First, due to the oscillating factors in Eq.~(\ref{eq:40})
only the states which
satisfy the coherency constraint
$m^{*^2} - m_{p}^{2} \lsim Q^{2}/R_{A}m_{p}$
can contribute to FSI. Second, the off-diagonal diffraction
amplitudes $f(i N\rightarrow j N)$ also become small
when the masses $m_{i}$ and $m_{j}$ differ strongly.

The above suppression of the heavy intermediate states makes
the theoretical predictions insensitive to the specific form
of the point-like initial ejectile wave function.
For instance, for Gaussian parameterization
used in the present paper
the missing momentum distribution must be
insensitive to the value of $C$ as long as
$C\gsim 1/Q^{2}\rho_{o}^{2}\,$,
where $\rho_{o}$ denotes the position of
the first node in the wave functions of the excited states
satisfying the coherency requirement.
We checked that in the region of $Q^{2}\lsim 40$ GeV$^{2}$, which we
discuss  in the present paper, our numerical
results are practically independent of the parameter $C$
for $C\gsim 0.1$.
It is worth noting that
the weak sensitivity of CT effects to the specific
choice of the point-like initial ejectile wave function
also vindicates neglecting the difference between the
CT effects
for the longitudinal and transverse spectral functions.

In conclusion of this section
one remark on the nonrelativistic
description of the $3q$ system is in order.
Of course, the nonrelativistic approach can not be
justified for the high excited states.
However, our numerical results show that due to the coherency
constraint the dominant role in the regime of the onset of CT
plays the first excitation of the proton. This,
in part, vindicates the use of the nonrelativistic model.
Still, we regard the nonrelativistic
quark-diquark model
only as a basis which allows us to obtain a realistic diffraction
scattering matrix, which is truly important from the
point of view of CCMST.

\section{Qualitative analysis of the incoherent FSI}
As was shown in section 2 an evaluation of the missing momentum
distribution with the \GG term included requires the information
on the $3q$-nucleon scattering matrix at arbitrary momentum transfer.
Despite the ensuing model-dependence of an analysis of the
inclusive $(e,e'p)$ reaction, certain
conclusions on the role of the
incoherent rescatterings can be reached
without specifying the explicit form of the scattering matrix.
Here we consider the simpler case of the integrated nuclear
transparency. For the case of the inclusive $(e,e'p)$ reaction we can
write
\beq
T_{A}^{inc}=
T_{A}^{exc}+\Delta T_{A}\,,
\label{eq:90}
\eeq
where
\beq
T_{A}^{exc}=
\int d^{3}\vec{r}\rho_{A}(\vec{r}\,)\Phi_{coh}(\vec{r},\vec{r}\,)\,,
\label{eq:91}
\eeq
is the transparency for exclusive
$(e,e'p)$ scattering when only the coherent rescatterings are allowed,
and the contribution of the
incoherent FSI is given by
\beq
\Delta T_{A}=
\int d^{3}\vec{r}\rho_{A}(\vec{r}\,)
[\Phi(\vec{r},\vec{r}\,)-\Phi_{coh}(\vec{r},\vec{r}\,)]\,.
\label{eq:92}
\eeq

To simplify the problem let us consider the two-channel model,
which involves only one resonance state $|p^{*}\rangle$.
We estimate $\Delta T_{A}$ expanding  the
evolution operator (\ref{eq:31})
up to first order in the \GG term. Notice that the one-fold incoherent
rescattering practically saturates the missing momentum distribution
in the region of $p_{m}\lsim 300$ MeV/c, which gives the dominant
contribution to $T_{A}^{inc}$ \cite{NSZ}. Also, we neglect the
off-diagonal coherent rescatterings. Then, making use of
Eqs.~(\ref{eq:22}),~(\ref{eq:31}),~(\ref{eq:35}),~(\ref{eq:92})
after some simple algebra we  get
\bea
\Delta T_{A}=
\frac{1}{16\pi^{2}}
\int
d^{2}\vec{q}\left [
|\langle p|\hat{f}(\vec{q}\,)|p\rangle |^{2} I_{pp}+
|\langle p|\hat{f}(\vec{q}\,)|p^{*}\rangle |^{2}
I_{p^{*}p^{*}}|C_{p^{*}}|^{2} \right.\nonumber\\
\left.+
2\mbox{Re}\langle p|\hat{f}(\vec{q}\,)|p^{*}\rangle
\langle p|\hat{f}(\vec{q}\,)|p\rangle^{*} I_{p^{*}p}C_{p^{*}}\right]
\,.\;\;\;\;\;\;\;\;\;\;\;\;\,
\label{eq:94}
\eea
Here
\beq
I_{pp}=
\int d^{2}\vec{b}_{1}dz_{1}\rho_{A}(\vec{b}_{1},z_{1})
\int\limits_{z_{1}}^{\infty} dz
\,n_{A}(\vec{b}_{1},z)
\exp\left[-\mbox{Re}\sigma_{pp}t(\vec{b}_{1},\infty,z_{1})\right]\,,
\label{eq:95}
\eeq
\bea
I_{p^{*}p^{*}}=
\int d^{2}\vec{b}_{1}dz_{1}\rho_{A}(\vec{b}_{1},z_{1})
\int\limits_{z_{1}}^{\infty} dz
n_{A}(\vec{b}_{1},z)\nonumber\;\;\;\;\;\;\;\;\;\;\;\;\;\;\;\;\\
\;\;\;\;\;\;\;\;\;\;\times
\exp\left[-\mbox{Re}\sigma_{pp}t(\vec{b}_{1},\infty,z)-
\mbox{Re}\sigma_{p^{*}p^{*}}t(\vec{b}_{1},z,z_{1})\right]\,,
\label{eq:96}
\eea
\bea
I_{p^{*}p}=
\int d^{2}\vec{b}_{1}dz_{1}\rho_{A}(\vec{b}_{1},z_{1})
\int\limits_{z_{1}}^{\infty} dz
\,n_{A}(\vec{b}_{1},z)
\exp(ik_{pp^{*}}z)\nonumber\;\;\;\;\;\;\;\;\;\;\;\;\;\\
\;\;\;\;\;\;\;\;\;\;\times
\exp\left[-\mbox{Re}\sigma_{pp}t(\vec{b}_{1},\infty,z)-
\frac{(\sigma_{p^{*}p^{*}}+\sigma_{pp}^{*})}{2}
t(\vec{b}_{1},z,z_{1})\right]\,.
\label{eq:97}
\eea

The diagonal and off-diagonal amplitudes in
Eq.~(\ref{eq:94}) are related through the CT sum
rule
\beq
\langle p|\hat{f}(\vec{q}\,)|p\rangle +
\langle p|\hat{f}(\vec{q}\,)|p^{*}\rangle
C_{p^{*}}=0\,.
\label{eq:99}
\eeq
Making use of (\ref{eq:99}) we can write
(\ref{eq:94}) as
\beq
\Delta T_{A}=
\sigma_{el}(pN)
\left [
I_{pp}+I_{p^{*}p^{*}}
-2\mbox{Re}I_{p^{*}p}\right]
\,.
\label{eq:94p}
\eeq
At low energy of the struck proton, when
$k_{pp^{*}}R_{A}\approx (m_{p^{*}}^{2}-m_{p}^{2})R_{A}/2\varepsilon\gg 1$,
the $p^{*}p$ interference term in Eq.~(\ref{eq:94p})
(the last term in the square brackets
in the right hand side of Eq. (\ref{eq:94p}))
becomes small and we get
\beq
\Delta T_{A}\approx
\sigma_{el}(pN)\left[I_{pp}+I_{p^{ * }p^{*}}\right]\,.
\label{eq:98}
\eeq
At high energy, when
$k_{pp^{*}}R_{A}\approx (m_{p^{*}}^{2}-m_{p}^{2})R_{A}/2\varepsilon\ll 1$,
the $p^{*}p$ interference term in Eq.~(\ref{eq:94p})
is not suppressed and will in part
cancel the contributions of the first two terms.
Thus, we see that, on the contrary to
the coherent
FSI, in the case of the incoherent FSI the CT effects
decrease nuclear transparency. Hence, a
conspiracy of the
CT effects related to the coherent and incoherent rescatterings
must take place in the integrated nuclear transparency
measured in inclusive $(e,e'p)$ reaction.
According to the qualitative consideration of section 2 and
the analysis \cite{NSZ},
the contribution of the incoherent rescatterings to the missing
momentum distribution for the most part comes from the region
of $p_{m\perp}\gsim 200-250$ MeV/c.
Evidently, the above discussed difference between
$\Delta T_{A}$	at low and high energy comes  namely from
this region of the missing momentum.
Hence, it is advantageous to perform experimental
measurements of the nuclear transparency
separately in the regions of
$p_{m\perp}\lsim 200$ MeV/c and $p_{m\perp} \gsim 250$ MeV/c.
At small momenta the CT signal is increasing the transparency,
while at large momenta the CT  will manifest itself through
decreasing the transparency.

Eq.~(\ref{eq:98}) demonstrates that,
on the contrary to the wide spread opinion,
in inclusive $(e,e'p)$ reaction the contribution of
the off-diagonal rescatterings survives at low energies.
To this effect,
the inclusive $(e,e'p)$ scattering
differs  drastically from exclusive $(e,e'p)$ reaction or elastic
hadron-nucleus scattering, where at low energies the predictions of
CCMST and the Glauber model are close to each other.
Of course, one should bear in mind that Eq.~(\ref{eq:98}) is
obtained in the idealized quark model, which ignores
the finite value of the resonance width, $\Gamma_{p^{*}}$.
Inclusion of the finite $\Gamma_{p^{*}}$ will lead
to a suppression of the second term in the right-hand side of
Eq.~(\ref{eq:98}) at sufficiently small energies
$\varepsilon \lsim \Gamma_{p^{*}}m_{p^{*}}l_{int}\sim 2$ GeV
(here $l_{int}\sim (\sigma_{tot}(pN)\langle n_{A}\rangle)^{-1}$
is the average interaction length of the $3q$ system in the nuclear
medium).
However, it is important that the scale of the
energy , where the finite-width effects
become strong, are
by the factor $\sim 3-5$ smaller than the energy scale
of the CT effects $\varepsilon\sim (m_{p^{*}}^{2}-m_{p}^{2})R_{A}/2$.
It means that there is a certain energy interval in which the CT
effects are still small but, none the less,
the Glauber model is not justified for
evaluation of the contribution to the missing momentum distribution
of the incoherent rescatterings.

It is appropriate here to
comment on the previous analyses of the $Q^{2}$-dependence of
the integrated nuclear transparency. In refs.
\cite{OnsetCT,BBK1,JK} the
calculations were performed making use of the optical
potential form of the FSI factor.
It means that the integrated transparency
of refs. \cite{OnsetCT,BBK1,JK} corresponds to
the exclusive $(e,e'p)$ reaction.
The authors of refs.
\cite{Kohama1,Kohama2} also used the optical potential
FSI factor. However, they replaced the total $pN$ cross section
by the inelastic $pN$ cross section, $\sigma_{in}(pN)$.
In the Glauber model the integrated transparency in the inclusive
$(e,e'p)$ scattering is indeed controlled by $\sigma_{in}(pN)$
(see discussion of this problem in \cite{NSZ}).
The analysis of the present paper makes clear that the CT effects
from the incoherent rescatterings
can not be described by a simple renormalization of the
diffraction scattering matrix in the equations obtained in
the optical potential approach. For this reason prescription
of refs. \cite{Kohama1,Kohama2} is not justified.

It is worth noting that the above analysis
indicates that in the case of the quasielastic $(p,2p)$ scattering
also a complicated interplay of the CT effects from  the coherent
and incoherent rescatterings may take place. Evidently, in this
process too the contribution of the off-diagonal incoherent
rescatterings will survive at low energies. In
$(p,2p)$ scattering the contribution of the incoherent rescatterings
appears to be considerably enhanced in a comparison with $(e,e'p)$
reaction \cite{dipsig4}. For this reason an analysis of
the CT effects in $(p,2p)$ scattering must include the incoherent
rescatterings, which were neglected in all previous works
on this problem.

\section{Numerical results for the exclusive
\lowercase{$(e,e'p)$} scattering}
In this section we present our numerical results
for the nuclear transparency in exclusive
$(e,e'p)$ scattering obtained making use of Eqs.~(\ref{eq:37})-(\ref{eq:40}).
We remind, that in the region of $p_{m}
\mathrel{\rlap{\lower4pt\hbox{\hskip1pt$\sim$}}
\raise1pt\hbox{$<$}} 150-200$ MeV/c,
where the contribution of the incoherent rescatterings becomes
small, our theoretical predictions may be compared directly
with experimental data for inclusive $(e,e'p)$ reaction.
The numerical calculations were
carried out for the target nuclei $^{16}O$ and $^{40}Ca$.
We used in our calculations the harmonic oscillator
shell wave functions.
The oscillator shell model frequency, $\omega_{osc}$, for
the two nuclei were adjusted to reproduce the experimental value of the
root-mean-square radius of the charge distribution,
$\langle r^{2}\rangle^{1/2}$. We used the values \cite{Atdata}
$\langle r^{2}\rangle^{1/2}=2.73$ fm for $^{16}O$, and
$\langle r^{2}\rangle^{1/2}=3.47$ fm for $^{40}Ca$,
which correspond to the oscillator
radius, $r_{osc}=(m_{p}\omega_{osc})^{-1/2}$, equal to
1.74 fm for $^ {16}O$ and  1.95 fm for $^{40}Ca$.
The difference between the charge distribution and the proton
nuclear density connected with the proton charge radius was
taken into account.
We checked that
our set of the harmonic oscillator shell wave functions
gives the charge density and SPMD
in the region of $p_{m}
\mathrel{\rlap{\lower4pt\hbox{\hskip1pt$\sim$}}
\raise1pt\hbox{$<$}}$(250-300)\,Mev/c, which
are practically indistinguishable from the results of more involved
Hartree-Fock calculations. Notice that in this momentum region
SPMD calculated in the harmonic oscillator shell model is
also close to the one obtained within
a many-body approach with realistic nucleon-nucleon potential
in ref. \cite{BCS}.
In our calculations we define the $pN$ cross section and
$\alpha_{pN}$ as mean values of these quantities for the
$pp$ and $pn$ scatterings. We borrowed the experimental data
on $pp$, $pn$ cross sections and $\alpha_{pp}$, $\alpha_{pn}$ from
the recent review \cite{Lehar}.

To illustrate the role of the off-diagonal rescatterings,
which are responsible for the CT effects,
we present a systematic comparison of the results obtained within
CCMST and the ones obtained in the Glauber model.
The number of the included resonance states and the
the multiplicity of the off-diagonal rescatterings used in
Eqs.~(\ref{eq:38})-(\ref{eq:40}) to obtain the curves corresponding to
CCMST were equal to 4 and 3, respectively.
 We checked that the contributions from higher
excitations and rescatterings with $\nu > 3$ are negligible
in the region of
$Q^{2}\mathrel{\rlap{\lower4pt\hbox{\hskip1pt$\sim$}}
 \raise1pt\hbox{$<$}} 40$ GeV$^{2}$ considered
in the present paper.

In Figs.~2,~3 we show the behavior of the nuclear transparency
versus	$p_{m,z}$ for the purely parallel kinematics at
$Q^{2}=5,\,10,\,20\,$ and $40$ GeV$^{2}$. Notice that the nuclear
transparency evaluated
even in the Glauber model without off-diagonal rescatterings
has sizeable
asymmetry about $p_{m,z}=0$ connected with nonzero $\alpha_{pN}$,
which was neglected in previous works.
Figs.~2,~3 demonstrate that the CT effect in the region of $Q^{2}
\mathrel{\rlap{\lower4pt\hbox{\hskip1pt$\sim$}} \raise1pt\hbox{$<$}}
 10$ GeV$^{2}$ is still small at $|p_{m,z}|
\mathrel{\rlap{\lower4pt\hbox{\hskip1pt$\sim$}}
\raise1pt\hbox{$<$}} 150$ MeV/c. However,
at $p_{m,z}\sim -250$ MeV/c it becomes sizeable and could be
observed in a high precision experiment. Our calculations show that
the situation is more favorable in the case of light nuclei.

The results for the nuclear transparency for
the transverse kinematics are presented in Figs.~4,~5.
We see that, as in the case of the parallel kinematics,
the distortion effects are considerable even in the Glauber model.
The CT effects are still small in the region of $Q^{2}
\mathrel{\rlap{\lower4pt\hbox{\hskip1pt$\sim$}}
\raise1pt\hbox{$<$}} 10$ GeV$^{2}$.
They become important only at $Q^{2}
 \mathrel{\rlap{\lower4pt\hbox{\hskip1pt$\sim$}} \raise1pt\hbox{$>$}} 20$
 GeV$^{2}$, especially at
large $p_{m\perp}$.

In Figs.~6,~7 we show our predictions for the integrated nuclear transparency.
In order to demonstrate the dependence of the
nuclear transparency on the choice of the kinematical domain $D$
in the definition (\ref{eq:2}), we calculated $T_{A}$ for
four different windows in the missing momenta.
One sees that the most steep rise of $T_{A}$ takes place
for the window containing negative values of the longitudinal
missing momentum.

We also calculated the integrated nuclear transparency
for the excitation of separate hole states in the target
nucleus for the kinematical domain $p_{m\perp}, |p_{m,z}|<200$ MeV/c.
The results are presented in Figs.~8,~9.
These figures show that CT effects are different for the
different shell states, being larger for the 1s state.
This fact is further illustrated by Figs.~10-13,
which show the missing momentum distribution for excitation of
the separate hole states for the
parallel and transverse kinematics
at $Q^{2}=5$ and 40 GeV$^{2}$.
Unfortunately, however, the experimental information
about deeply bound hole states are most difficult to
extract, since their strength is fragmented over
a wide range missing energy.

The relative CT effect of different excitations of
the $3q$ ejectile state is demonstrated in Fig.~10, where  we plot the
integrated
nuclear transparency for the window $p_{m\perp}, |p_{m,z}|<200$ MeV/c
calculated for the number of included intermediate states, $n$,
equals 1,~2,~3,~4. We see that in the region $Q^{2}
\mathrel{\rlap{\lower4pt\hbox{\hskip1pt$\sim$}} \raise1pt\hbox{$<$}} 40$
GeV$^{2}$ the FSI effects are practically saturated for $n=3$.
At
$Q^{2} \mathrel{\rlap{\lower4pt\hbox{\hskip1pt$\sim$}}
\raise1pt\hbox{$<$}} 20$ GeV$^{2}$ in the CCMST formalism it is
sufficient to take into account
only the first excitation of the proton.

Our numerical results show that the experimental observation
of the CT phenomenon at
$Q^{2} \mathrel{\rlap{\lower4pt\hbox{\hskip1pt$\sim$}}
 \raise1pt\hbox{$<$}}  20$ GeV$^{2}$ is a delicate
problem. For this reason it is important to understand
how large are the uncertainties of the theoretical predictions
for the contribution of the off-diagonal rescatterings.
The least reliable ingredient
is the reggeon part of the $3q$-nucleon amplitudes.
To estimate the corresponding uncertainties we studied
the sensitivity of the results to the choice of the reggeon
parameters
$\alpha_{1}$ and $\alpha_{2}$ in (\ref{reggeon}), which
control the diagonal
and off-diagonal matrix elements, respectively.
Variation of $\alpha_{1}$ practically does not change
the results. However, the dependence on $\alpha_{2}$ is not negligible.
The suppression of the off-diagonal $3q$-nucleon
amplitudes in comparison with the diagonal ones is expected to be
more strong for the reggeon exchange than for the Pomeron one.
For this reason we chose for the upper bound of $|\alpha_{2}|$
the value 0.5, which is about the maximum value of $|\alpha_{pN}|$.
The effect of variation of $\alpha_{2}$ in the range
( -0.5,0.5) for integrated nuclear transparency
in the window $p_{m\perp}, |p_{m,z}|<200$ MeV/c is illustrated
in Fig.~15. As one can see, the positive values of $\alpha_{2}$
decrease the nuclear transparency, and can to a certain extent
obscure the
CT effects at $Q^{2} \mathrel{\rlap{\lower4pt\hbox{\hskip1pt$\sim$}}
 \raise1pt\hbox{$<$}}  20$ GeV$^{2}$. In Figs.~16,~17 we
demonstrate the effect of variation of $\alpha_{2}$
for the unintegrated nuclear
transparency for the parallel kinematics. It is seen that,
the gross features of the $p_{m,z}$-dependence of the CT effect
are stable with respect to variation of $\alpha_{2}$.
In the transverse kinematics the variation
of $\alpha_{2}$ yields only the overall renormalization of the
nuclear transparency. Thus, as far as the F-B asymmetry of the
nuclear transparency is concerned, the uncertainties of
the reggeon amplitudes can not change the situation considerably.
However, we are bound to conclude
that at $Q^{2} \mathrel{\rlap{\lower4pt\hbox{\hskip1pt$\sim$}}
 \raise1pt\hbox{$<$}} 20$ GeV$^{2}$ the real situation may be
more complicated for the observation of CT
through $Q^{2}$-dependence of the integrated nuclear transparency.

In addition to the above discussion on the uncertainties of
the theoretical predictions it is also appropriate to comment on the
off-shell effects, which are neglected in our analysis.
A successful observation of the missing momentum dependence of the
CT effects is only possible provided that the uncertainties of
the missing momentum distribution extracted from the measured
cross section of $(e,e'p)$ scattering are small in a comparison
with the theoretically calculated
contribution of the off-diagonal rescatterings.
The determination of the missing momentum distribution
includes the division of the experimental $(e,e'p)$ cross section
by the half off-shell $ep$ cross section.
As a consequence, ambiguities in $\sigma_{ep}$ lead to unavoidable
uncertainties in the extracted missing momentum distribution.
In order to estimate these uncertainties
we compared
the off-shell $ep$ cross sections evaluated under different
prescriptions discussed in ref. \cite{Forest}. We found that
the typical off-shell ambiguities are $
\mathrel{\rlap{\lower4pt\hbox{\hskip1pt$\sim$}}
\raise1pt\hbox{$<$}} 5-10$\%
in the
kinematical region considered in the present paper.
 Such uncertainties are not big  enough to obscure the CT effects
at $Q^{2}\sim 10$ GeV$^{2}$ for the parallel kinematics,
where CT effects increase the ratio
$T_{A}(-p_{m,z},p_{m\perp}=0)/T_{A}(p_{m,z},p_{m\perp}=0)$
at $p_{m,z}\sim 250$ MeV/c
by the factor $\sim 2$ (see Figs.~2,~3).
At higher values of $Q^{2}$
($ \mathrel{\rlap{\lower4pt\hbox{\hskip1pt$\sim$}}
 \raise1pt\hbox{$>$}} 20$ GeV$^{2}$)
the off-shell uncertainties can be neglected
both for the parallel and transverse kinematics.

\section{Conclusions}
We have studied the missing momentum dependence of the CT effects
in $(e,e'p)$ scattering in the kinematical region
of $p_{m}
\mathrel{\rlap{\lower4pt\hbox{\hskip1pt$\sim$}}
\raise1pt\hbox{$<$}}
 250$ MeV/c and $Q^{2}
\mathrel{\rlap{\lower4pt\hbox{\hskip1pt$\sim$}}
\raise1pt\hbox{$<$}}
 40$ GeV$^{2}$.
To perform such an analysis we developed a formalism based on the
Glauber-Gribov multiple scattering theory.
In our calculations we describe the target nucleus in the
independent particle shell model.
The formalism of CCMST was presented in the form which includes
both the coherent and incoherent rescatterings.
The coherent rescatterings describe FSI in exclusive
$(e,e'p)$ reaction, when only the one-hole excitations of the target
nucleus are allowed, while inclusion of both the coherent and incoherent
rescatterings corresponds to inclusive experimental conditions
involving all final states of the residual nucleus.
The CT effects related to the incoherent rescattering were not
considered in previous works.

We performed a qualitative analysis of the off-diagonal
incoherent rescatterings making use of the two-channel model.
Our important observation is that, on the contrary to the
case of coherent FSI,  the contribution of the
off-diagonal incoherent rescatterings does not vanish at small
energies of the struck proton. For this reason, even at low $Q^{2}$,
 in the case of
the inclusive reaction the Glauber model becomes unreliable for
treatment of FSI in the region of large
missing momenta, where the incoherent rescatterings dominate.
We demonstrated that CT leads to a decrease of the contribution
of the incoherent rescatterings and a conspiracy of the CT
effects from the coherent and incoherent rescatterings may
take place in measurement of the integrated nuclear transparency.
We argue that this phenomenon may be important in $(p,2p)$
scattering, where the contribution of the incoherent rescatterings
are enhanced in a comparison with $(e,e'p)$ reaction.

The numerical calculations of the present paper
were carried out for the exclusive
$(e,e'p)$ reaction. In the region of $p_{m}
\mathrel{\rlap{\lower4pt\hbox{\hskip1pt$\sim$}}
\raise1pt\hbox{$<$}} 150-200$ MeV/c,
where the effect of the incoherent FSI becomes small, our predictions
can be compared with the experimental data obtained in inclusive
$(e,e'p)$ reaction.
Our calculations show that at $Q^{2}
\mathrel{\rlap{\lower4pt\hbox{\hskip1pt$\sim$}}
\raise1pt\hbox{$<$}} 5$ GeV$^{2}$ the CT
effects are still small in the whole missing momentum region
considered in the present paper.
At $Q^{2}\sim 10$ GeV$^{2}$ we find
a considerable CT effect only in the case of the parallel
kinematics in the region of $p_{m,z}\sim -(200-250)$ MeV/c.
The CT increases the ratio
$T_{A}(-p_{m,z},p_{m\perp}=0)/T_{A}(p_{m,z},p_{m\perp}=0)$
at $p_{m,z}\sim 250$ MeV/c
by the factor $\sim 2$.
This effect is stronger for the light target nuclei. It
could be observed in a high precision experiment.
Our calculations show that the developed CT regime
starts with $Q^{2}\sim 40$ GeV$^{2}$, where the CT effects
change the missing momentum distribution drastically.

We studied for the first time the impact of the reggeon
exchanges on the CT effects.
It was found that in the region of
$Q^{2} \mathrel{\rlap{\lower4pt\hbox{\hskip1pt$\sim$}}
 \raise1pt\hbox{$<$}} 40$
GeV$^{2}$ the effect of the diagonal $3q$-nucleon reggeon amplitudes
is practically saturated by the elastic $pN\rightarrow pN$
amplitude. However, the off-diagonal
resonance-nucleon reggeon amplitudes may be important.
We found that
the onset of the CT regime for integrated nuclear transparency
may be delayed if the reggeon exchanges generates the positive
value of the  Re/Im ratio for the off-diagonal amplitudes.
None the less the CT effect in the F-B asymmetry is
insensitive to the reggeon amplitudes.

For the first time we presented a detailed analysis of the
convergence
of the CCMST series in the number of the included resonance states.
Our results indicate that at $Q^{2}
\mathrel{\rlap{\lower4pt\hbox{\hskip1pt$\sim$}}
\raise1pt\hbox{$<$}} 40$ GeV$^{2}$
the first 2-3 excited states practically saturate the contribution
of the off-diagonal rescatterings in exclusive $(e,e'p)$ scattering.

The results obtained for the integrated nuclear transparency
show that the observation of
CT through the $Q^{2}$-dependence of the integrated nuclear transparency
is hardly possible at $Q^{2}
\mathrel{\rlap{\lower4pt\hbox{\hskip1pt$\sim$}}
 \raise1pt\hbox{$<$}} 10$ GeV$^{2}$. However,
extension of the kinematical region up to $Q^{2}\sim 40$ GeV$^{2}$
could provide the observation of CT if the small-size $3q$ configurations
actually dominate in hard $ep$ scattering.
\vspace{.2cm}\\
\acknowledgements

 This work was partly supported by the
 Grant N9S000 from the International Science Foundation and
 the INTAS grant 93-239.
AAU acknowledges Prof. A. Zichichi and ICSC- World Laboratory
for financial
support.
BGZ wishes to gratefully acknowledge the hospitality of the
Interdisciplinary Laboratory of SISSA and the Institut
f\"ur Kernphysik, KFA, J\"ulich.


\pagebreak
\begin{figure}
\caption{The typical diagrams contributing to the operator
$\hat{U}(\vec{r}_{1},\vec{r}_{1}^{\,'})$ describing the
evolution of the density matrix of $3q$ ejectile state within CCMST:
(a) the diagram without the interaction between the two
trajectories outgoing from $\vec{r}_{1}$ and $\vec{r}_{1}^{\,'}$,
(b) the diagram containing the interaction between the trajectories
generated by the \GG term in Eq.~(23). The dotted lines
attached to the straight-line trajectory originating from
$\vec{r}_{1} (\vec{r}_{1}^{\,'}$) denote a profile function
$\hat\Gamma(\vec{b}_{j}-\vec{b}_{1},z_{j}-z_{1}) (\hat\Gamma^{*}
(\vec{b}_{j}-\vec{b}_{1}^{\,'},z_{j}-z_{1}^{'})$).}

\caption{ Nuclear transparency in exclusive $^{16}O(e,e'p)$ scattering
in parallel kinematics $p_{m\perp}=0$
calculated within CCMST (solid curve) and in the Glauber model
(dotted curve).}

\caption{ The same as Fig.~2, but for $^{40}Ca(e,e'p)$ scattering.}

\caption{ Nuclear transparency in exclusive $^{16}O(e,e'p)$ scattering
in transverse kinematics $p_{m,z}=0$
calculated within CCMST (solid curve) and in the Glauber model
(dotted curve).}

\caption{ The same as Fig.~4, but for $^{40}Ca(e,e'p)$ scattering.}

\caption{ The $Q^{2}$-dependence of nuclear transparency for
exclusive $^{16}O(e,e'p)$ scattering at different windows $D$ in the
transverse and longitudinal missing
momentum
obtained  within CCMST (solid curve) and in the Glauber model
(dotted curve).}

\caption{ The same as Fig.~6, but for $^{40}Ca(e,e'p)$ scattering.}

\caption{ The $Q^{2}$-dependence of nuclear transparency for
exclusive $^{16}O(e,e'p)$ scattering for excitations of
the separate hole states
at the kinematical window $p_{m\perp},~|p_{m,z}|< 200$ MeV/c
obtained  within CCMST (solid curve) and in the Glauber model
(dotted curve).}

\caption{ The same as Fig.~8, but for $^{40}Ca(e,e'p)$ scattering.}

\caption{ The missing momentum distribution for $^{16}O(e,e'p)$
scattering in parallel kinematics $p_{m\perp}=0$ for the separate
shells calculated within CCMST (solid curve) and in the
Glauber model (dotted curve). The dot-dashed curve shows the
SPMD.}

\caption{ The same as Fig.~10, but for $^{40}Ca(e,e'p)$
scattering.}

\caption{ The same as Fig.~10, but for transverse kinematics.}

\caption{ The same as Fig.~11, but for transverse kinematics.}

\caption{The convergence of CCMST expectation for
nuclear transparency
in exclusive $^{16}O(e,e'p)$ and $^{40}Ca(e,e'p)$ scattering
for the missing
momentum window $p_{m\perp},~|p_{m,z}|<200$ Mev/c
with respect to the number
of the $3q$ states included: $n=1$ (solid curve),
$n=2$ (long-dashed curve), $n=3$ (dot-dashed curve),
$n=4$ (dotted curve).}

\caption{ Nuclear transparency for exclusive $^{16}O(e,e'p)$ and
$^{40}Ca(e,e'p)$ scattering for the missing momentum window
$p_{m\perp},~|p_{m,z}|<200$ Mev/c calculated within
CCMST with different sets of the reggeon parameters:
$\alpha_{2}=0$ ( solid curve), $\alpha_{2}=-0.5$ ( long-dashed curve),
$\alpha_{2}=0.5$ ( short-dashed curve),
in all the cases $\alpha_{1}=1$. The predictions of the
Glauber model are shown by the dotted curve.}

\caption{ The $p_{m,z}$-dependence of the nuclear
transparency for exclusive $^{16}O(e,e'p)$ scattering.
The legend of curves is the same as in Fig.~15.}

\caption{ The same as Fig.~16, but for $^{40}Ca(e,e'p)$
scattering.}

\end{figure}
\end{document}